\documentclass[twoside,twocolumn,9pt]{article}
\usepackage{extsizes}
\usepackage[super,sort&compress,comma]{natbib} 
\usepackage[version=3]{mhchem}
\usepackage{soul}
\usepackage[left=1.5cm, right=1.5cm, top=1.785cm, bottom=2.0cm]{geometry}
\usepackage{balance}
\usepackage{mathptmx}
\usepackage{sectsty}
\usepackage{graphicx} 
\usepackage{lastpage}
\usepackage[format=plain,justification=justified,singlelinecheck=false,font={stretch=1.125,small,sf},labelfont=bf,labelsep=space]{caption}
\usepackage{float}
\usepackage{fancyhdr}
\usepackage{fnpos}
\usepackage[english]{babel}
\addto{\captionsenglish}{%
  
}
\usepackage{array}
\usepackage{droidsans}
\usepackage{charter}
\usepackage[T1]{fontenc}
\usepackage[usenames,dvipsnames]{xcolor}
\usepackage{setspace}
\usepackage[compact]{titlesec}
\usepackage{hyperref}
\newcommand{\CC}{[$-$C$_2-$ANT$-$C$_2-$]$_n$}
\newcommand{\CAu}{[$-$Au$-$C$_2-$ANT$-$C$_2-$]$_n$}
\usepackage{epstopdf}%This line makes .eps figures into .pdf - please comment out if not required.

\definecolor{cream}{RGB}{222,217,201}

\begin{document}

\pagestyle{fancy}
\thispagestyle{plain}
\fancypagestyle{plain}{
%%%HEADER%%%
\renewcommand{\headrulewidth}{0pt}
}
%%%END OF HEADER%%%

%%%PAGE SETUP - Please do not change any commands within this section%%%
\makeFNbottom
\makeatletter
\renewcommand\LARGE{\@setfontsize\LARGE{15pt}{17}}
\renewcommand\Large{\@setfontsize\Large{12pt}{14}}
\renewcommand\large{\@setfontsize\large{10pt}{12}}
\renewcommand\footnotesize{\@setfontsize\footnotesize{7pt}{10}}
\makeatother

\renewcommand{\thefootnote}{\fnsymbol{footnote}}
\renewcommand\footnoterule{\vspace*{1pt}% 
\color{cream}\hrule width 3.5in height 0.4pt \color{black}\vspace*{5pt}} 
\setcounter{secnumdepth}{5}

\makeatletter 
\renewcommand\@biblabel[1]{#1}            
\renewcommand\@makefntext[1]% 
{\noindent\makebox[0pt][r]{\@thefnmark\,}#1}
\makeatother 
\renewcommand{\figurename}{\small{Fig.}~}
\sectionfont{\sffamily\Large}
\subsectionfont{\normalsize}
\subsubsectionfont{\bf}
\setstretch{1.125} %In particular, please do not alter this line.
\setlength{\skip\footins}{0.8cm}
\setlength{\footnotesep}{0.25cm}
\setlength{\jot}{10pt}
\titlespacing*{\section}{0pt}{4pt}{4pt}
\titlespacing*{\subsection}{0pt}{15pt}{1pt}
%%%END OF PAGE SETUP%%%

%%%FOOTER%%%
\fancyfoot{}

\fancyfoot[RO]{\footnotesize{\sffamily{1--\pageref{LastPage} ~\textbar  \hspace{2pt}\thepage}}}
\fancyfoot[LE]{\footnotesize{\sffamily{\thepage~\textbar\hspace{4.65cm} 1--\pageref{LastPage}}}}
\fancyhead{}
\renewcommand{\headrulewidth}{0pt} 
\renewcommand{\footrulewidth}{0pt}
\setlength{\arrayrulewidth}{1pt}
\setlength{\columnsep}{6.5mm}
\setlength\bibsep{1pt}
%%%END OF FOOTER%%%

%%%FIGURE SETUP - please do not change any commands within this section%%%
\makeatletter 
\newlength{\figrulesep} 
\setlength{\figrulesep}{0.5\textfloatsep} 

\newcommand{\topfigrule}{\vspace*{-1pt}% 
\noindent{\color{cream}\rule[-\figrulesep]{\columnwidth}{1.5pt}} }

\newcommand{\botfigrule}{\vspace*{-2pt}% 
\noindent{\color{cream}\rule[\figrulesep]{\columnwidth}{1.5pt}} }

\newcommand{\dblfigrule}{\vspace*{-1pt}% 
\noindent{\color{cream}\rule[-\figrulesep]{\textwidth}{1.5pt}} }

\makeatother
%%%END OF FIGURE SETUP%%%

%%%TITLE, AUTHORS AND ABSTRACT%%%
\twocolumn[
  \begin{@twocolumnfalse}
\par
\vspace{1em}
\sffamily
\begin{tabular}{m{0.5cm} p{15.5cm} }

& \noindent\LARGE{\textbf{Steric hindrance in the on-surface synthesis of diethynyl-linked anthracene polymers}}
%Structural Robustness Induced by Steric Hindrance of Assembled Diacetylene Linked Anthracene Oligomers}}
%acetylene e oligomers non vengono mai riportati nel testo
\\%Article title goes here instead of the text "This is the title"
\vspace{0.9cm} & \vspace{0.3cm} \\

 & \noindent\large{Simona Achilli,\textit{$^{b\ast}$} and Francesco Tumino,\textit{$^{a\ddag}$} and Andi Rabia,\textit{$^{a}$} and Alessio Orbelli Biroli,\textit{$^{c,d}$} and Andrea Li Bassi,\textit{$^{a}$} and Alberto Bossi,\textit{$^{c}$} and Nicola Manini,\textit{$^{b}$} and Giovanni Onida,\textit{$^{b}$} and Guido Fratesi,\textit{$^{b}$} and Carlo Spartaco Casari,\textit{$^{a}$}} \\%Author names go here instead of "Full name", etc.

\\

  & \noindent\normalsize{Hybrid sp-sp$^2$ structures can be efficiently obtained on metal substrates via on-surface synthesis. The choice of both the precursor and of the substrate impacts on the effectiveness of the process and the stability of the formed structures. Here we demonstrate that using anthracene-based molecules as precursor, the formation on Au(111) of polymers hosting sp carbon chains is affected by the steric hindrance between aromatic groups. In particular, by scanning tunneling microscopy and density functional theory calculations we show that the de-metalation of organometallic structures induces a lateral separation of adjacent polymers preventing the formation of ordered domains.} \\%The abstrast goes here instead of the text "The abstract should be..."

\end{tabular}

 \end{@twocolumnfalse} \vspace{0.9cm}

]
%%%END OF TITLE, AUTHORS AND ABSTRACT%%%

%%%FONT SETUP - please do not change any commands within this section
\renewcommand*\rmdefault{bch}\normalfont\upshape
\rmfamily
\section*{}
\vspace{-1cm}

%%%FOOTNOTES%%%

\footnotetext{\textit{$^{a}$~Department of Energy, Politecnico di Milano, via Ponzio 34, Milano, Italy.}}
\footnotetext{\textit{$^{b}$~ETSF and Dipartimento di Fisica “Aldo
Pontremoli”, Università degli Studi di Milano, via Celoria 16, Milano, Italy}}
\footnotetext{\textit{$^{\ast}$}simona.achilli@unimi.it}
\footnotetext{\textit{$^{c}$ Istituto di Scienze e Tecnologie Chimiche “G. Natta”, Consiglio Nazionale delle Ricerche (CNR-SCITEC),via Golgi 19 - 20133 Milano; PST via G. Fantoli 16/15 - 20138 Milano; SmartMatLab Centre, via Golgi 19 - 20133 Milano, Italy}}
\footnotetext{\textit{$^{d}$ Dipartimento di Chimica, Università di Pavia, via Taramelli 12 - 27100 Pavia, Italy}}
\footnotetext{\textit{$^{\ddag}$}francesco.tumino@polimi.it}

%Please use \dag to cite the ESI in the main text of the article.
%If you article does not have ESI please remove the the \dag symbol from the title and the footnotetext below.
%\footnotetext{\dag~Electronic Supplementary Information (ESI) available: [details of any supplementary information available should be included here]. See DOI: 10.1039/cXCP00000x/}
%additional addresses can be cited as above using the lower-case letters, c, d, e... If all authors are from the same address, no letter is required

%%%END OF FOOTNOTES%%%

%%%MAIN TEXT%%%%

\section{Introduction}
Carbon exhibits unique capability in forming a tremendous variety of nanostructures of different size, dimensionality and properties.\cite{Hirsch_NatMat_2010} After the advent of graphene, research was devoted to the production of graphene nanoribbons, with the target of opening up a bandgap by exploiting lateral confinement effects.\cite{son2006energy}
Bottom up fabrication \cite{Fasel2010, Chen2013-GNR} turned out to be a viable alternative to etching graphene edges,\cite{Datta08}, unzipping of carbon nanotubes, \cite{Jiao09} or exploit lithographic methods. \cite{Achilli14} 
To achieve this goal, the synthesis process requires an extremely fine control of the structures down to the atomic scale.
The on-surface synthesis approach allows the fabrication of carbon nanostructures by promoting chemical reactions directly on a metal surface, and enables the use of in-situ surface science characterization techniques.\cite{SHEN201777, Ruffieux2016489, steiner2017hierarchical, Clair2019-OSS} In particular, scanning tunneling microscopy (STM) and atomic force microscopy gives access to the atomic-scale and molecular imaging of the structure and arrangement of the sample with unprecedented capability, opening
a way to the
controlled fabrication
of atomically-precise nanostructures.\cite{Shu_NatComm_2018}

By playing with $\pi$-conjugated
molecular precursors used as the building units in the on-surface synthesis process, a number of different systems has been developed so far, including one-dimensional (1D) and two-dimensional (2D) polymeric structures supported on metal surfaces. \cite{zuzak2020, Galeotti-NatMat2020, Sedona}
More recently such approach has been extended to peculiar carbon structures involving combined
sp-sp$^2$ hybridization by exploiting the dehalogenative homocoupling reaction of precursor molecules with ethynyl groups.\cite{Sun_dehalogenative, Yu2020, rabia2019scanning, Rabia_nanomat}
1D linear structures (i.e.\ sp-sp$^2$ carbon polymers) and 2D sp-sp$^2$ carbon networks (i.e.\ graphdiynes) have been synthesized on metal surfaces and investigated.

The on-surface synthesis process has been unveiled in detail as well as the structural, electronic and vibrational properties of these systems.
Such properties have been shown to be affected 
by the metal substrate 
mainly
through charge-transfer processes and 
structural distortions imposed to adapt the carbon network to the
periodicity of the underlying metal surface.\cite{rabia2019scanning, Rabia_nanomat, Achilli-GDY}

Nonetheless,
the choice of the precursor 
was
demonstrated to be 
the
primary aspect in the control of the on-surface synthesis process.\cite{Xu17,Wang21}
The rational design of the 
precursor
molecules allows one --for example-- to exploit their side functionality to control the order and symmetry of the polymers, and the steric hindrance of chemical groups to steer the molecular assembly.\cite{Chem-Rev, Rev-SSR, Xiang19} 
Topologically non-trivial properties have been unveiled on $\pi$-conjugated polymers made by laterally assembling acene monomers,
as a function of the
number of the aromatic rings.
In particular moving from anthracene- to pentacene-based polymers, a
change
from insulating to metallic behavior is expected, due to a topological phase transition.\cite{Cirera}
Most of these 
investigations
focus mainly on the single wire or polymer as well as on the single molecular bond,\cite{Kawai18, Cirera}
while
paying less attention to the polymer-polymer interaction and to the overall self-assembly on the metal surface.
Due to the complex on-surface synthesis process involving different steps from the dehalogenation to the homocoupling, the self-assembling behavior is non trivial, and it can be affected by different aspects which are at present unexplored.

In the present paper, we
report a combined theoretical and experimental investigation of anthracene-based polymers synthesized on Au(111) and in situ investigated by STM.

Through the deposition of 9,10-di(bromoethynyl)anthracene on Au(111) followed by thermal annealing we obtain densely packed organometallic polymers (i.e. with Au adatoms bridging the ethynyl groups) as first stage and homocoupled diethynyl-linked anthracene polymers (\CC) at high temperature. Upon the release of the Au bridging atoms we assist to a sideward distancing and bending of adjacent polymers resulting in more disordered structure. 
We demonstrate through density functional theory calculations that such a transition - from close-packed domains to disordered polymers- is due to the polymer-polymer interaction and in particular to the steric hindrance of lateral aromatic rings of adjacent molecules.
This effect, caused by the shortening of the homocoupled diethynyl chain due to the desorption of the Au atom is also responsible of a different registry of anthracene units of neighboring chains.

Moreover the calculations reveal that the interaction with the substrate determines the metallic character of both the organometallic and homocoupled polymers.

We show that the overall behavior of an on-surface synthesized system results from a not trivial interplay between intermolecular effects and substrate interaction. A full understanding of these mechanisms is mandatory to engineer novel systems with \textit{ad hoc} structure and electronic properties.

\section{Results and discussion}

\subsection{STM experiments}

\begin{figure*}[ht]
	\centering
	\includegraphics[width=.7\textwidth]{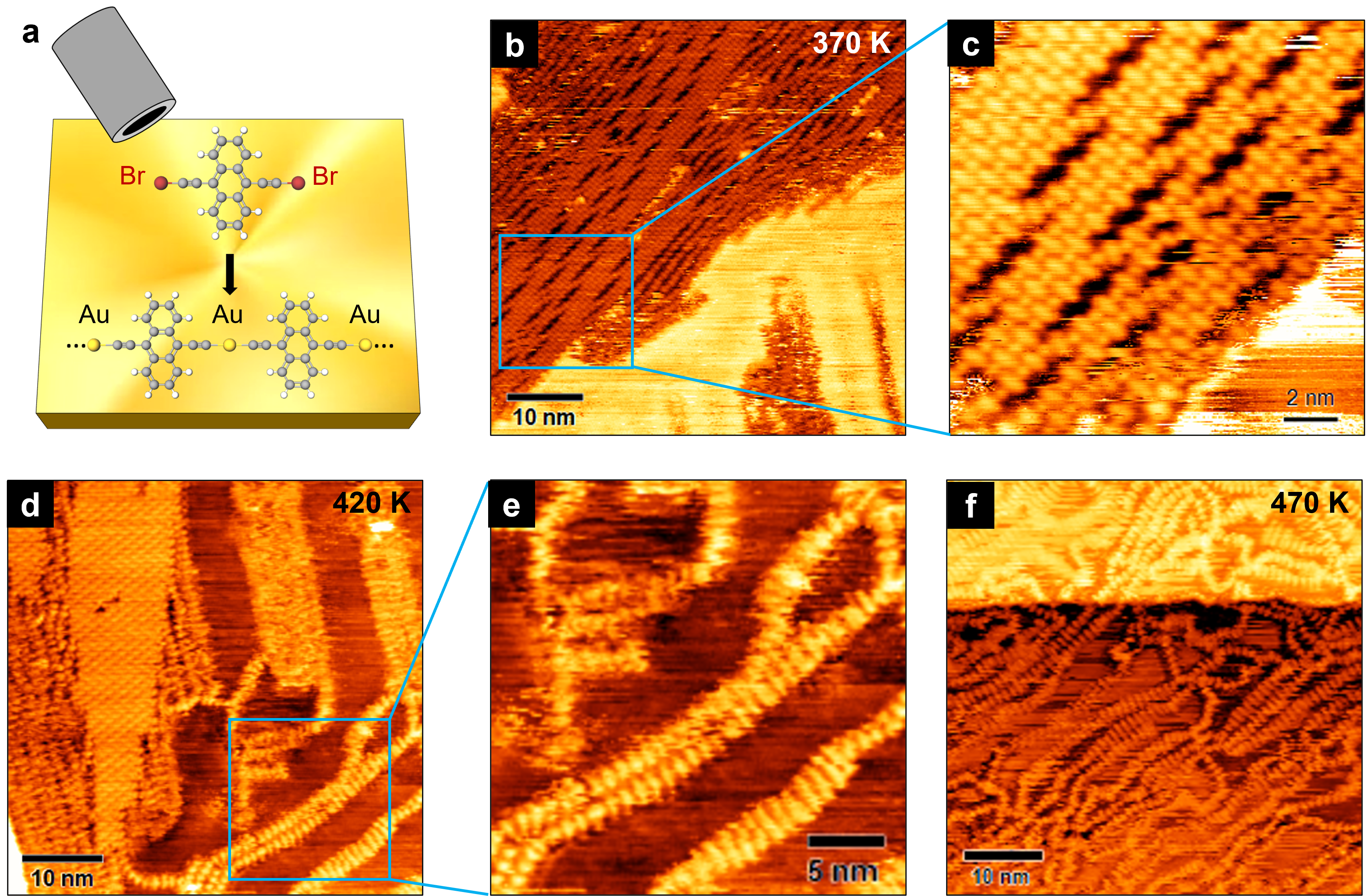}
	\caption{(a) Schematic representation of the synthesis process: precursor molecules are evaporated onto the Au(111) surface (top half of the image), which catalyzes the debromination and the coupling of monomers into an organometallic polymer. (b) 50$\times$50 nm$^2$ STM image acquired after annealing at 370 K, showing the formation of ordered domains of packed organometallic chains (0.5 V, 0.2 nA). (c) Close-up of the region inside the box in (b). (d) STM image acquired after annealing at 420 K, showing the coexistence of compact domains and isolated linear structures (1 V, 0.2 nA). The latter are shown at higher resolution in (e). (f) STM image acquired after annealing at 470 K, showing the disordered arrangement of isolated chains (1.7 V, 0.3 nA).
	}
	\label{STM_morphology}
\end{figure*}

\begin{figure}[ht]
	\centering
	\includegraphics[width=.5\textwidth]{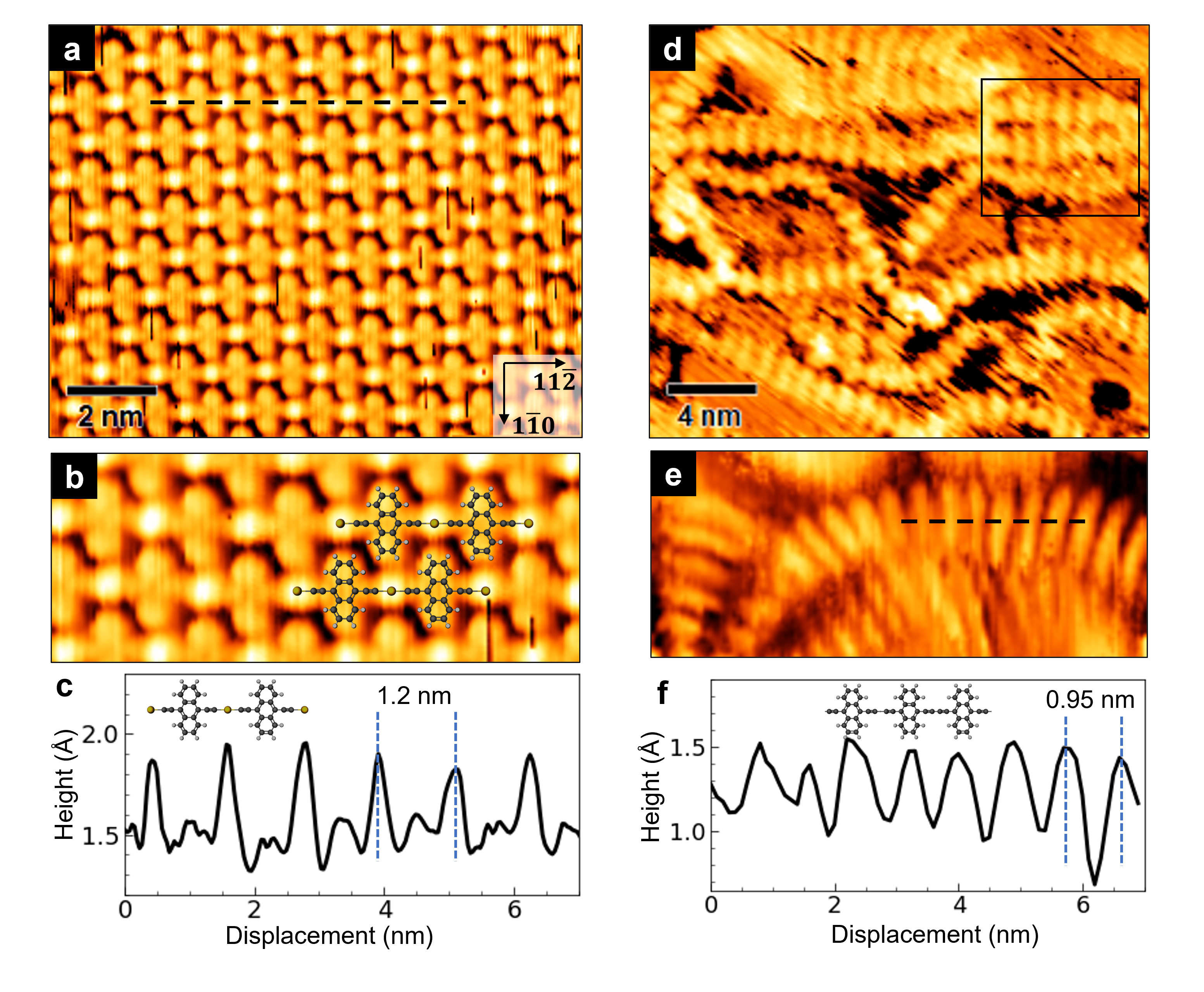}
	\caption{(a) Molecular-scale STM image showing the linear organometallic chains packed in a well-oriented domain (0.5 V, 0.2 nA). (b) Close-up on adjacent organometallic chains, whose structure is highlighted by the superimposed ball and stick model.
	(c) 
	Height
	profile along the chain 
	axis,
	specifically along
	the dashed line in (a).
	(d) STM image showing disordered chains after annealing at 470~K (0.7 V, 0.2 nA).
	The black frame highlights adjacent chains whose anthracene units are nearly aligned in a side-by-side arrangement.
	(e) Molecular-scale STM image of a 
	single homocoupled chain (1.7 V, 0.3 nA).
	(f) 
	Height profile along the homocoupled chain in (d).
	}
	\label{STM_structure}
\end{figure}

We have achieved the on-surface synthesis of 1D ethynyl-linked anthracene polymers
on Au(111) through the deposition of the precursor molecule shown in Fig.~\ref{STM_morphology}a. 
The molecular assembly into longer chains occurs on Au(111) at room temperature (RT) via a three-fold process\cite{Rabia_nanomat}: (i) adsorption and diffusion on the surface (ii) debromination reaction (cleavage of C-Br bonds), and (iii) coupling of ethynyl chains to gold adatoms.
The resulting structures are 1D organometallic polymers, consisting in gold-mediated ethynyl chains linked to anthracene (\CAu). 
We imaged these structures after a mild annealing of the sample at 370 K, which favors the desorption of possible impurities coming from the evaporation and allows us to acquire more stable STM images.
Fig.~\ref{STM_morphology}b shows the organometallic polymers at a relatively large scale.
They tend to arrange
themselves
in compact domains with many chains running almost parallel to each other.
The packing of adjacent chains in ordered domains is interposed by voids (Fig. \ref{STM_morphology}c) which may occur periodically, suggesting the possibility that the packing of the linear structures could be hindered by internal forces.

By increasing the annealing temperature to 420~K (Fig. \ref{STM_morphology}d), the compact domains coexist with more isolated linear structures (Fig.~\ref{STM_morphology}e). As we will discuss below, these chains have a different structure compared to those forming the ordered domains. 
Also, the fuzzier imaging suggests that these chains have a higher mobility on the Au surface. 
After annealing at 470~K (Fig. \ref{STM_morphology}f), we observe only disordered chains, suggesting that a thermally activated mechanism induces the closely-packed organometallic polymers to gradually convert into linear structures randomly arranged on the Au surface.

By means of high-resolution STM measurements,
here we focus on
the structure of the chains packed in compact domains and the isolated ones observed in the 470~K - annealed sample.

The STM image in Fig.~\ref{STM_structure}a shows a portion of a well-ordered domain, observed in the sample annealed at 420 K.
The organometallic chains are oriented along the $[11\bar 2]$ direction of the substrate.
The bright spots alternating with the anthracene units along a single chain are
produced by
the gold atoms which mediate the coupling (see the model structure in Fig.~\ref{STM_structure}b).
Evidently, the well-oriented domains of organometallic chains are packed in such a way that the anthracene units
come close
to the gold adatoms of the 
adjacent
linear molecules.
The periodicity along the chain is $(12\pm 0.3)$~\AA{}
(Fig. \ref{STM_structure}c);
the 
average
distance 
between adjacent chains along the $[1\bar 10]$ direction is $8.4 \pm 0.3$~\AA.

Fig.~\ref{STM_structure}d shows disordered chains observed after annealing at 470 K.
There is no evidence of the gold spot between 
%the 
successive [$-$C$_2-$ANT$-$C$_2-$] monomers
(Fig.~\ref{STM_structure}e), a sign that gold adatoms have been released and 
the coupling between ethynyl terminal groups has occurred.
This transformation is accompanied by a shortening of the periodicity along the chains, which reduces to $(9.5\pm 0.03)$~\AA\ (Fig. \ref{STM_structure}f).
Also, the anthracene units of nearby chains are no longer in the alternating configuration of organometallic chains, but rather they tend to arrange 
themselves
in a side-by-side configuration (see framed region in Fig.~\ref{STM_structure}d) and, more importantly, they are spaced 
farther apart than in the ordered structure. 
A minimum distance of about 17~\AA~ can be estimated when two or more chains are close together, such as in the framed region of Fig.~\ref{STM_structure}d).

The tighter in-chain packing of the anthracene units produces a looser lateral packing of the polymeric chains, due to steric hindrance.

\subsection{Theoretical structural models}

In order to 
obtain
an insight into the formation mechanism of homocoupled \CC\ 
structures we perform density functional theory (DFT) calculations aimed to characterize the different stages of the on-surface-synthesis process.

\begin{figure*}[ht]
  \centering 
	\includegraphics[height=9cm]{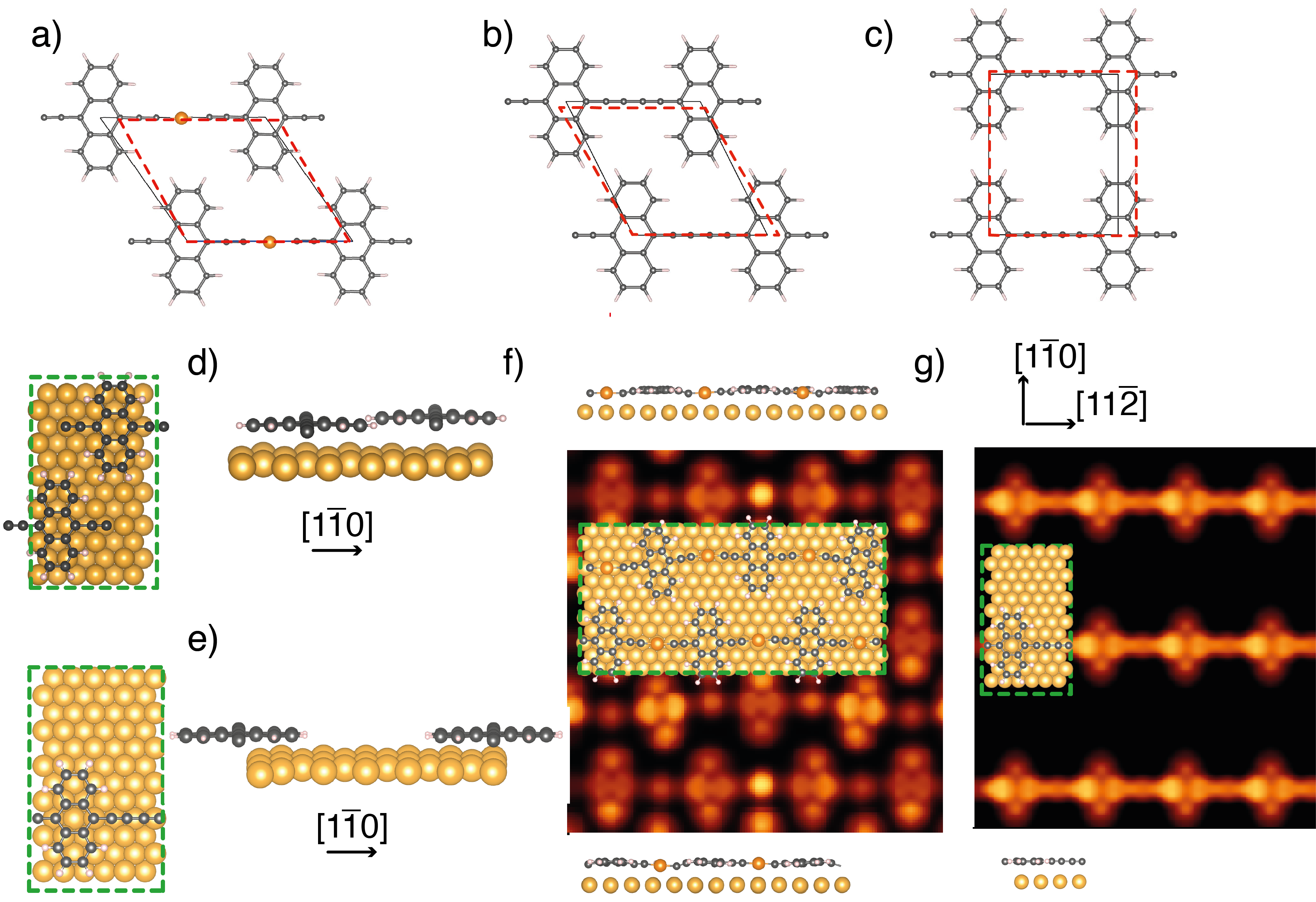}
	\caption{Structural model for the freestanding \CAu\ (a) and \CC\ in the staggered (b) and aligned configuration (c). Solid gray: the relaxed unit cell.
	Dashed red: the best fitting Au-substrate-compatible periodic cell.
	(d) and (e):
    the two possible models for \CC\ on Au(111) and the related structural relaxation of the anthracene units.
    (f) and (g): the simulated STM images and the relaxed structures of \CAu\ and \CC\ on Au(111).
    }
	\label{Fig-model}
\end{figure*}

As a preliminary step we investigate, via structural relaxation, the optimal periodicity and stacking of the linear organometallic compounds in an
ideal
freestanding 
2D array 
configuration, before and after the release of the gold linking atom.
The resulting structural models are reported in Fig.~\ref{Fig-model}a-c.
This figure additionally reports
the unit cell adapted to 
the closest-matching periodicity
of Au(111)
(red dashed)
compared to
the relaxed unit cell of the freestanding model system (black solid).

The relaxed freestanding \CAu~forms organometallic chains with a 
spacing
of 12.09 \AA~ between
successive
anthracene units, in agreement with the experimental observation, and a
spacing of 9.19 \AA~ between parallel linear structures (horizontal and vertical size of the black cell of Fig.~\ref{Fig-model}a, respectively).

The matching with the gold substrate (theoretical equilibrium lattice constant $a_{Au}=4.21$~\AA) is imposed by considering a supercell formed by 7 Au unit cell vectors along the $[11\bar 2]$ direction ($a_{11\bar2}=a_{Au}\times\sqrt{3/2}=5.16$~\AA)  and 6 unit cell vectors along the [1$\bar1$0] direction ($a_{1\bar10}=a_{Au}/\sqrt{2}=2.98$~\AA) (green rectangle in Fig.~\ref{Fig-model}f) that accommodates two adjacent chains with three \CAu\ monomer units each.
Reporting this choice to the hexagonal unit-cell of the freestanding polymer (red dashed in Fig. \ref{Fig-model}a), the unit-cell lattice vectors (modulus $7/3 a_{11\bar2}$ and $2\sqrt{3}a_{1\bar10}$, respectively) are slightly compressed with respect to the freestanding equilibrium model and the angle between them is reduced from 125$^\circ$ to 120$^\circ$. Consequently, the separation between successive anthracene units and between parallel polymers is reduced (-0.47\% along the [$11\bar2$] direction and -1.6\% along the [$1\bar10$], respectively). 
The energy cost for matching amounts to 3 meV/monomer for the freestanding system.

For \CC\ we consider two possible configurations: the staggered one (Fig. \ref{Fig-model}b) which is the ground state for 
the 2D
freestanding 
crystal of this polymer,
and the aligned one (Fig.~\ref{Fig-model}c) which has been chosen as representative of the arrangement observed in the STM images.
In the first one the relaxed unit-cell vectors (black in Fig.~\ref{Fig-model}b) form an angle of $116.7^\circ$ and measure 9.52~\AA\ in the horizontal direction and 10.59~\AA\ in the rotated one. 
In the aligned configuration, the equilibrium unit cell is a rectangle with vector
lengths
9.53~\AA\ and 11.52~\AA\ in the horizontal and vertical direction, respectively (black in Fig.~\ref{Fig-model}c).
It is worth noting that the most relevant aspect of the this configuration is that the separation between the linear chains, and consequently between 
adjacent 
anthracene units, is larger than in the staggered one.
Small deviations from the orthogonality of the cell vectors, as observed in the experiment, would not change the considerations reported below.   
The freestanding staggered stacking is 35~meV/monomer
more stable than the aligned one.

The matching with the underlying Au substrate introduces 
in both cases a 
significant
strain along the $[11\bar 2]$ direction (8.4\% and 8.3\% tensile strain in the staggered and aligned cases, respectively) which is necessary to fit the structure with 2 Au(111) cell vectors.
The energy cost associated to such expansion of the polymer amounts to $\sim 1.9$~eV/monomer.
To obtain a mismatch lower than 1\% also in this case, a unit cell including 6 monomer units fitting 11 Au(111) cell vectors would be required.

Along the $[1\bar 10]$ direction the staggered configuration matches to a 3 Au lattice spacing with a 5.5\% compression.  This is associated to a small energy of 8~meV/monomer as no covalent bond is compressed and adjacent chains can interpenetrate to some extent.

Differently, the aligned configuration is nearly commensurate with
4
Au lattice spacings along the $[1\bar 10]$ direction (3.3\% strain, 1~meV/monomer
energy cost).
The adapted unit cells are reported in Fig.~\ref{Fig-model} b and c (red dashed).
We verified that a further expansion of the cell vector along
$[1\bar 10]$ in the aligned model has a reasonably small cost, of the order of 30 meV/monomer
in the freestanding system. Such a small deformation energy would not prevent the matching with a larger Au unit cell in this direction when the system is grown on the substrate, in agreement with the STM images that show a separation between the polymers compatible with 6 Au(111) unit-cell vectors along the $[1\bar 10]$ direction (see the frame in Fig.~\ref{STM_structure}d).

As a further test we 
perform an additional
calculation for a 2D network of freestanding  anthracene-based monomers (not linked by the terminal ethynyl groups), arranged analogously as in Fig.~\ref{Fig-model}b, where the
spacing between molecular rows in the vertical direction (referring to the figure) is fixed 
to the value
observed in the experiments for \CAu\ (8.4~\AA\ along $[1\bar 10]$ direction) and varying the side-by-side separation of anthracene units.

We find an increase of energy for approaching anthracene units, with an energy cost of 0.5 eV when the periodicity along $[1 1\bar 2]$
is 
9.50~\AA, as in the experiment.

On the basis of these results and of the information extracted from the STM images, we can infer that the formation of \CC\ polymer on the Au(111) surface is made difficult by the steric hindrance of the anthracene units depending on the reduction of their distance induced by the release of the gold adatom.

This steric effect prevents the \CC\ polymers from packing in ordered domains after their formation.
Indeed 
a certain degree of lateral order
takes place only in
less densely packed regions
of the surface (Fig.~\ref{STM_structure}d) where the separation of the linear chains is possible, leading to 
partly ordered
structures resembling the aligned arrangement
of Fig.~\ref{Fig-model}c.

To verify this assumption we 
execute the DFT simulations of
\CC\ adsorbed on Au(111) surface, 
in
both the staggered and the aligned configuration.
For both, we
use the same supercell, based on
a $2 \times 6$ rectangular cell of Au(111) (green in Fig. \ref{Fig-model}d and e).
This cell includes 2 monomer units in the staggered configuration,
and 1 in the aligned one.

The polymer separation in the aligned case fits the periodicity observed in the STM images resulting larger than in the equilibrium freestanding system.

By comparing the total DFT energy per \CC~ unit on Au(111) in the two models  (Fig.~\ref{Fig-model}d and e), we obtain
an energy gain of 0.47~eV/monomer for the aligned configuration 
relative to the staggered one, in agreement with the result found for isolated freestanding anthracene molecules%anthracene-based monomers
at nearly the same distance (see above).

The calculated adsorption energy, obtained as:
$$E_{ads}=(E_{tot}-E_{mon}-E_{sub})/n_{mon},$$
where $E_{tot}$, $E_{mon}$ and $E_{sub})$ are the (per monomer) energy of the adsorbed system, of the freestanding relaxed
polymer,
and of the substrate $2\times 6$ gold
substrate,
respectively, 
amounts to
$-8.32$~eV for the \CAu~ on Au(111).
For the \CC~ polymer on Au(111) it is 
$-4.32$~eV/monomer
for the staggered configuration and $-4.79$~eV/monomer
for the aligned one, confirming that the \CC\ polymers are more weakly bound to the surface and the inter-molecular distance is larger
compared to the equilibrium freestanding case.

\subsection{Local structure and electronic properties}
Fig.~\ref{Fig-model}f-g report the top view and side view of the relaxed geometries of \CAu\ and \CC\ on Au(111) and the simulated STM images for filled states (integration in the $0.5$~eV-wide interval immediately below the Fermi energy) at a constant distance of 2~\AA\ 
above
the
average height of the
molecular layer. 
The structural relaxation performed for \CAu\ leads to a partial in-plane bending of the linear chains which is driven by the preference of the Au atoms in the chain to sit at
a hollow position (see ball and stick model in Fig. \ref{Fig-model}f).
Such bending cannot be 
detected
in the high-resolution STM images reported in Fig.~\ref{STM_structure}a-b.

This feature might in fact
be partially due to the forced matching to the underlying Au substrate periodicity imposed in the calculation, which is also affected by the 3\% discrepancy between the experimental and 
the DFT-simulated
lattice spacing of gold.
It is quite possible
that a longer supercell in
the $[11\bar 2]$ direction
compatible with a smaller strain
of the polymer
might produce smaller or even no
bending.

The side view of the two linear fragments of parallel chains belonging to the unit cell (top and bottom of Fig. \ref{Fig-model}f) shows an overall modulation of the linear structures, with anthracene units lying farther from the surface (distances ranging from 2.84~\AA~to 2.92~\AA)
than the diethynyl chains
that are pulled down
toward the surface due to the small distance ($\Delta$z=2.42~\AA)
between the gold adatom
and the gold surface underneath.
The STM simulation displays bright spots in correspondence of gold adatoms in the chain, in agreement with the experiments, and bright regions in correspondence of the external rings of anthracene.% questo è ridondante and of the lateral sp$^2$ carbon atoms of the central ring.

The two models considered for \CC\ on Au(111) show relevant differences in the relaxed geometry of the anthracene units: in the staggered configuration (Fig.~\ref{Fig-model}d) the steric hindrance induces a bending of the external aromatic rings of one anthracene unit in the cell (side view in Fig.~\ref{Fig-model}d).

In contrast,
in the aligned configuration the molecules preserve the planar geometries (see the side view in Fig.~\ref{Fig-model}e),
thus confirming the absence of steric hindrance.
Moreover, the average height of the
the polymers above
the substrate top layer is larger in the staggered configuration (2.77~\AA) than in the aligned one (2.71~\AA).

We have performed the STM simulation 
for
the model with aligned molecules,
with
the same setup as in Fig.~\ref{Fig-model}f.
In agreement with experiment, the bright
spot at the center of the diethynyl chain disappears, due to the absence of the gold
adatom.
Both the chain and the anthracene units are imaged with
similar brightness.

\begin{figure*}[ht]
  \centering 
	\includegraphics[height=7cm]{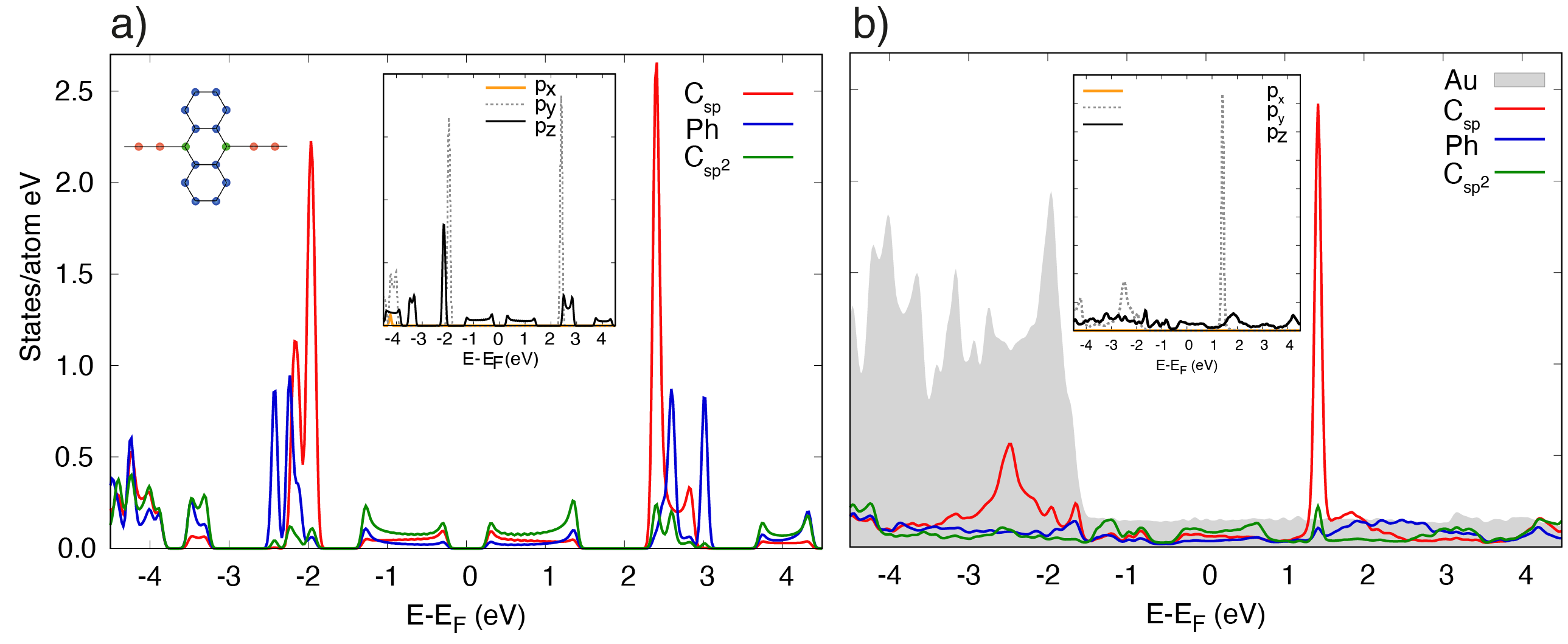}
	\caption{Projected density of states of (a) freestanding \CC, and (b) \CC/Au(111) in the aligned configuration. 
	The contributions of different groups of atoms in the polymer, namely sp carbon atoms in the chain (red), sp$^2$ carbon atoms linking the chain to the anthracene unit (green) and external phenyl rings (blue) are resolved (see also the color code in the inset).
	Grey shadow: density of states projected on the Au substrate.
	Insets:	The same density of states projected over the individual components of the p orbitals of the sp
	carbon atoms in the ethynyl chain.
	$x$ indicates the $[1\bar1 0]$ direction, $y$ is the $[11\bar2]$ direction, and $z$ is the direction perpendicular to the surface. 
	Each individual PDOS is normalized to the number of atoms of the relevant kind.
	}
	\label{Fig-dos}
\end{figure*}

This result can be rationalized in terms of
the computed
projected density of states (PDOS) of \CC/Au(111), reported in Fig.~\ref{Fig-dos}.
Below the Fermi level 
the PDOS exhibits
similar contributions coming from the lateral aromatic rings, ethynyl chain and sp$^2$ carbon atoms at the edge of anthracene, as color-coded in the sketched molecule.

The insets
report a direction-resolution of the projected states on the ethynyl chain: features at the Fermi level have
a mainly
p$_z$ character;
p$_x$ states are mostly out of the energy range considered;
p$_y$ states, orthogonal to the chain direction, 
exhibit
two peaked features at $-2.5$ and $1.2$~eV that originate from the 
$\pi$-bonds
formed between sp carbon atoms and are 
relatively weakly
affected by the hybridization with the substrate.\cite{ma11122556}
A comparison with the PDOS of the freestanding 
polymer
evidences that the states at the Fermi level of \CC/Au(111) originate from a shift of the unoccupied DOS of the freestanding molecule
to lower energy:
this shift is 
associated
to a mild charge transfer to the 
polymer,
which
we evaluate in
0.12~electrons/monomer.

In particular, the largest amount of charge (0.10~e/monomer) is transferred to the anthracene unit while sp chain acquires only 0.02~electrons.

Accordingly, we find a similar depletion of charge in Au surface layer ($-0.005$~e/Au, $-0.11$~e per surface layer).

The most relevant effect of the interaction between the polymer and the substrate is that the polymer acquires a metallic character, despite a hint of the energy gap is still visible just below the Fermi level, although reduced to 0.2~eV 
compared
to the 0.4~eV of the freestanding 
polymer.

Hybridization with the d states of the substrate takes place via the p$_z$ and p$_y$ states of the sp chain below $-1.5$~eV and via the states of the aromatic ring in the same energy range which have p$_z$ character.

\begin{figure*}[ht]
  \centering 
	\includegraphics[height=7cm]{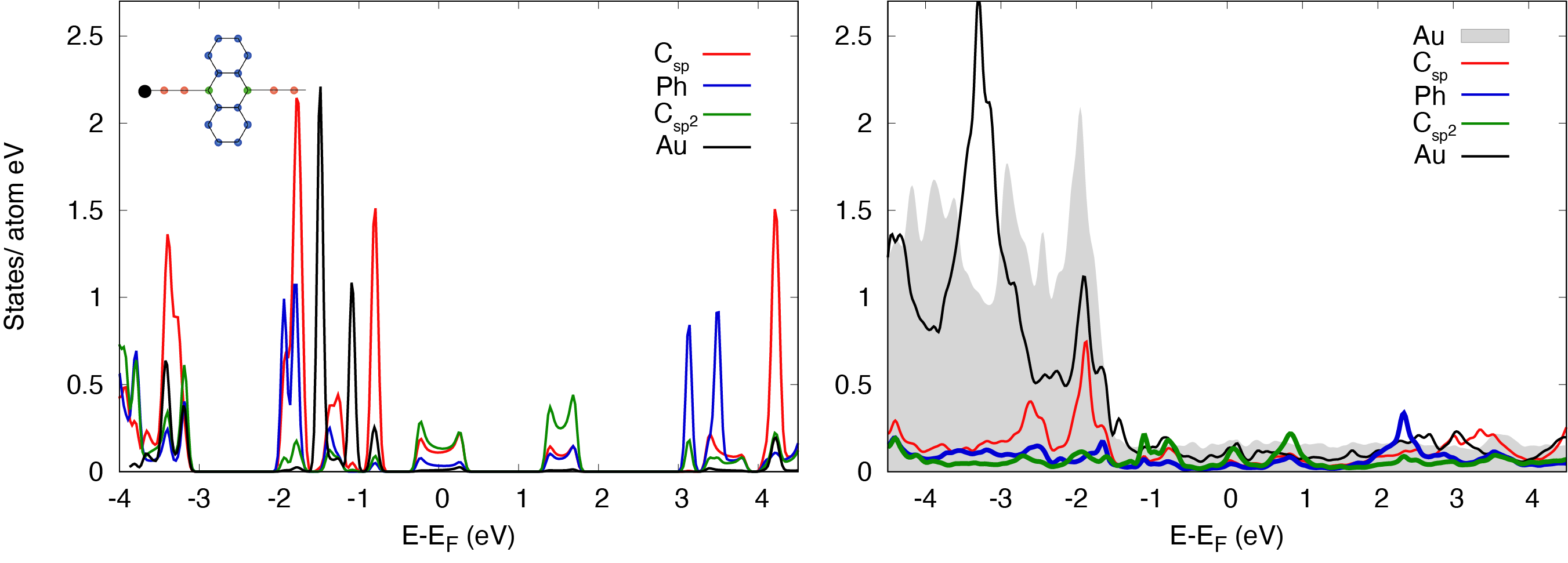}
	\caption{Projected density of states of (a) freestanding \CAu, and (b) \CAu/Au(111). 
	The contributions of different groups of atoms in the polymer, namely sp carbon atom in the ethynyl chain (red), sp$^2$ carbon atoms linking the chain to the anthracene unit (green), Au linking atom (black) and external phenyl rings (blue) are resolved (see also the color code in the inset).
	Grey shadow: density of states projected on the Au substrate.
	Each individual PDOS is normalized to the number of atoms of the relevant kind. The PDOS of the Au linking atom has been divided for a factor 5.
	}
	\label{dosCAu}
\end{figure*}

The interaction between the organometallic \CAu\ and the substrate is stronger, due to the linking Au atom.
The charge transfer to the molecule amounts to 0.25 electrons/monomer,
mainly directed to the anthracene and the diethynyl chain.
On the contrary,
the Au adatom in the chain transfers 0.13 electrons to the surface layer.
The PDOS's of \CAu, reported in Fig. \ref{dosCAu}, show the metallic character of the freestanding \CAu\ system which is mainly due to a charge transfer from sp carbon atoms to the bridging Au resulting into a shift of the HOMO states at at the Fermi level. A shrink of the pristine HOMO-LUMO bands is also observed resulting in a larger gap just above the Fermi level.
When the system is adsorbed on Au(111) the strong hybridization with the substrate produces a deformation of the features and the disappearence of any reminiscence of the original gap of \CC.
Indeed,
the PDOS of the Au adatom in the chain strongly overlaps with the substrate, giving rise to peaked structures in correspondence 
to
the d bands of the Au surface and a broad continuum of states in the energy range of the $s$ band of the substrate.

\section{Experimental}
\subsection{Synthesis of 9,10-di(bromoethynyl)anthracene}

All reagents and solvents were purchased from Merck and used without additional purification. $9,10-$diethynylanthracene was synthesized as previously reported in the literature.\cite{synthexp}
$^1$H and $^{13}$C 
nuclear magnetic resonance
(NMR) spectra were recorded by using a Bruker Avance DRX$-400$ or a Bruker Avance DRX$-300$ spectrometer in CDCl$_3$ (Cambridge Isotope Laboratories, Inc.) as solvent.
Mass spectra (MS) were obtained by means of an Advion expression CMS L01 instrument with an atmospheric-pressure chemical ionization source (APCI).

65.5~mg of 9,10-diethynylanthracene (0.289 $\mu$ mol)
were dissolved
in 6 mL of acetone. 
Under stirring 123.8~mg of N-bromosuccinimide (1.2 equiv.) and 12.0~mg of AgNO$_3$
(0.06 equiv.) were added, then the stirring
was maintained 
for 4 hours in the dark.
The solvent was removed in vacuum and the crude product was purified by flash column chromatography (SiO$_2$, eluent: $n$-hexane
gradient to 
$n$-hexane/CH$_2$Cl$_2$ 9:1).
46.4~mg of pure product
were obtained
(41.8\% yield).
$^1$H
NMR (400~MHz, CDCl$_3$, 20 °C): $\delta  = 8.63$ (m, 4H), 7.66 (m, 4H); 
$^{13}$C
NMR (300~MHz, CDCl$_3$,
20 °C): $\delta = 132.61$, 127.23, 126.77, 118.10, 77.06, 62.07.
MS (APCI$^+$): m/z (\%): 384.6 (100) [M$+$H]$^+$.

\subsection{On-surface synthesis and STM}
All experiments were conducted in an UHV system (base pressure $10^{-11}$ mbar) composed of two interconnected chambers for sample preparation and STM characterization. 
Au(111)/mica substrates (Mateck) were cleaned by cycles of Ar$^+$ sputtering (1 keV) and annealing at 700 K. 

The precursor in powder form was deposited on Au(111) at RT from an effusion cell (Dr. Eberl--MBE Komponenten) heated at 370 K.  
STM measurements were acquired at RT with an Omicron microscope, using homemade electrochemically etched W tips. 

Typical measurement parameters were in the range 0.5--2 V for bias voltage and 0.1--0.3 nA for set-point current (the specific values for the reported STM images are stated in the captions).

\subsection{Theory}
We perform DFT calculations with the generalized gradient approximation (GGA) in the PBE form
\cite{PBE}, and including
van der Waals interactions between the organic overlayer and the substrate via a DFT-D2
Grimme potential \cite{Grimme}.
We use the approach implemented in the SIESTA code \cite{Sole02} that relies on norm-conserving pseudopotentials and an atomic-orbitals
basis set. 
The simulated STM images are obtained in the Tersoff-Hamann approach \cite{Tersoff}.

We adopt a double-zeta basis set with polarization orbitals and a mesh-cutoff of 400~Ry for the kinetic-energy value that sets the real-space grid.

We relax the organic layers and the first substrate layer until the forces reach the tolerance value of 0.04~eV~\AA$^{-1}$.
We use a $4 \times 4$ sampling of the Brillouin zone for the freestanding polymers; for the calculations on Au we use a $2 \times 6$ sampling for \CAu\ and a $7 \times 6$ sampling for \CC.

Along the $z$ direction the slab includes three layers of Au substrate, the molecular overlayer and $\sim 85$~\AA\ of vacuum.

\section{Conclusions}

We demonstrate here the possibility to obtain hybrid sp-sp$^2$ carbon structures via on-surface synthesis reaction on Au(111) substrate by starting from a 9,10-di(bromoethynyl)anthracene.
The final product of the process are polymeric chains where the anthracene units are
alternating
with the ethynyl chains.
The STM images  
indicate
that the on-surface polymerization reaction proceeds through an organometallic polymer intermediate state arranged in compact and well-aligned rows.
However, the elimination of the gold adatoms leads to poorly-ordered assemblies of the purely organic \CC\ polymer.

Theoretical calculations performed for different structural
models, for both freestanding and Au-adsorbed polymers allow us to identify the steric hindrance between anthracene units as the main reason for the separation between the chains leading to the disordered arrangement.
The repulsion between the electronic clouds of adjacent molecules arises due to the shrinking of the polymer
upon the release of the gold adatom, combined with the periodicity constraints induced by the interaction with the substrate.
These results demonstrate the influence of the choice of the precursor on the properties of the final structures, confirming 
an interplay of
both the molecule-molecule and molecule-substrate interaction 
in the on-surface synthesis process.

\section*{Author Contributions}
Conceptualization, SA, AOB, GF, CSC;
%Data curation,
%Formal Analysis,
%Funding acquisition,
Investigation, SA, FT, AR, GF;
%Methodology,
%Project administration,
Resources, AOB, AB, CSC;
%Software,
%Supervision,
%Validation,
Visualization, SA, FT, AR;
Writing – original draft, SA, AR;
Writing – review \& editing, SA, FT, AR, AOB, ALB, AB, NM, GO, GF, CSC.
All authors discussed the research while this was performed, and have approved the final version of the manuscript.

\section*{Conflicts of interest}
There are no conflicts to declare.

\section*{Acknowledgements}
F.T., A.R., A.L.B. and C.S.C.\ acknowledge funding from the European Research Council (ERC) under the European Union’s Horizon 2020 research and innovation program ERC Consolidator Grant (ERC CoG2016 EspLORE grant agreement no.\ 724610, website: www.esplore.polimi.it).
S.A., G.F., N.M., and G.O.\ acknowledge CINECA for the use supercomputing facilities under the agreement with UNITECH-INDACO and Barcelona Supercomputing Center for the access to the Nasertic cluster of the Red Espa\~nola de Supercomputacion (grant FI-2021-2-0047).
N.M.\ acknowledges support from the grant PRIN2017 UTFROM of the Italian Ministry of University and Research.

%%%END OF MAIN TEXT%%%

\balance


%%%REFERENCES%%%
\bibliography{biblio} %You need to replace "rsc" on this line with the name of your .bib file

\providecommand*{\mcitethebibliography}{\thebibliography}
\csname @ifundefined\endcsname{endmcitethebibliography}
{\let\endmcitethebibliography\endthebibliography}{}
\begin{mcitethebibliography}{33}
\providecommand*{\natexlab}[1]{#1}
\providecommand*{\mciteSetBstSublistMode}[1]{}
\providecommand*{\mciteSetBstMaxWidthForm}[2]{}
\providecommand*{\mciteBstWouldAddEndPuncttrue}
  {\def\EndOfBibitem{\unskip.}}
\providecommand*{\mciteBstWouldAddEndPunctfalse}
  {\let\EndOfBibitem\relax}
\providecommand*{\mciteSetBstMidEndSepPunct}[3]{}
\providecommand*{\mciteSetBstSublistLabelBeginEnd}[3]{}
\providecommand*{\EndOfBibitem}{}
\mciteSetBstSublistMode{f}
\mciteSetBstMaxWidthForm{subitem}
{(\emph{\alph{mcitesubitemcount}})}
\mciteSetBstSublistLabelBeginEnd{\mcitemaxwidthsubitemform\space}
{\relax}{\relax}

\bibitem[Hirsch({2010})]{Hirsch_NatMat_2010}
A.~Hirsch, \emph{{Nat. Mater.}}, {2010}, \textbf{{9}}, {868--871}\relax
\mciteBstWouldAddEndPuncttrue
\mciteSetBstMidEndSepPunct{\mcitedefaultmidpunct}
{\mcitedefaultendpunct}{\mcitedefaultseppunct}\relax
\EndOfBibitem
\bibitem[Son \emph{et~al.}(2006)Son, Cohen, and Louie]{son2006energy}
Y.-W. Son, M.~L. Cohen and S.~G. Louie, \emph{Phys. Re. Lett.}, 2006,
  \textbf{97}, 216803\relax
\mciteBstWouldAddEndPuncttrue
\mciteSetBstMidEndSepPunct{\mcitedefaultmidpunct}
{\mcitedefaultendpunct}{\mcitedefaultseppunct}\relax
\EndOfBibitem
\bibitem[Jinming~Cai \emph{et~al.}(2010)Jinming~Cai, Ruffieux, Jaafar, Bieri,
  Braun, Blankenburg, Muoth, Seitsonen, Saleh, Feng, M\"ullen, and
  Fasel]{Fasel2010}
J.~Jinming~Cai, P.~Ruffieux, R.~Jaafar, M.~Bieri, T.~Braun, S.~Blankenburg,
  M.~Muoth, A.~P. Seitsonen, M.~Saleh, X.~Feng, K.~M\"ullen and R.~Fasel,
  \emph{Nature}, 2010, \textbf{466}, 470--473\relax
\mciteBstWouldAddEndPuncttrue
\mciteSetBstMidEndSepPunct{\mcitedefaultmidpunct}
{\mcitedefaultendpunct}{\mcitedefaultseppunct}\relax
\EndOfBibitem
\bibitem[Chen \emph{et~al.}(2013)Chen, de~Oteyza, Pedramrazi, Chen, Fischer,
  and Crommie]{Chen2013-GNR}
Y.-C. Chen, D.~G. de~Oteyza, Z.~Pedramrazi, C.~Chen, F.~R. Fischer and M.~F.
  Crommie, \emph{ACS Nano}, 2013, \textbf{7}, 6123--6128\relax
\mciteBstWouldAddEndPuncttrue
\mciteSetBstMidEndSepPunct{\mcitedefaultmidpunct}
{\mcitedefaultendpunct}{\mcitedefaultseppunct}\relax
\EndOfBibitem
\bibitem[Datta \emph{et~al.}(2008)Datta, Strachan, Khamis, and
  Johnson]{Datta08}
S.~S. Datta, D.~R. Strachan, S.~M. Khamis and A.-T.-C. Johnson, \emph{Nano
  Lett.}, 2008, \textbf{8}, 1912--1915\relax
\mciteBstWouldAddEndPuncttrue
\mciteSetBstMidEndSepPunct{\mcitedefaultmidpunct}
{\mcitedefaultendpunct}{\mcitedefaultseppunct}\relax
\EndOfBibitem
\bibitem[Jiao \emph{et~al.}(2009)Jiao, Zhang, Wang, Diankov, and Dai]{Jiao09}
L.~Y. Jiao, L.~Zhang, X.~R. Wang, G.~Diankov and H.~J. Dai, \emph{Nature},
  2009, \textbf{458}, 877--880\relax
\mciteBstWouldAddEndPuncttrue
\mciteSetBstMidEndSepPunct{\mcitedefaultmidpunct}
{\mcitedefaultendpunct}{\mcitedefaultseppunct}\relax
\EndOfBibitem
\bibitem[Achilli \emph{et~al.}({2014})Achilli, Tantardini, and
  Martinazzo]{Achilli14}
S.~Achilli, G.~F. Tantardini and R.~Martinazzo, \emph{{Phys. Chem. Chem.
  Phys.}}, {2014}, \textbf{{16}}, {17610--17616}\relax
\mciteBstWouldAddEndPuncttrue
\mciteSetBstMidEndSepPunct{\mcitedefaultmidpunct}
{\mcitedefaultendpunct}{\mcitedefaultseppunct}\relax
\EndOfBibitem
\bibitem[Shen \emph{et~al.}(2017)Shen, Gao, and Fuchs]{SHEN201777}
Q.~Shen, H.-Y. Gao and H.~Fuchs, \emph{Nano Today}, 2017, \textbf{13},
  77--96\relax
\mciteBstWouldAddEndPuncttrue
\mciteSetBstMidEndSepPunct{\mcitedefaultmidpunct}
{\mcitedefaultendpunct}{\mcitedefaultseppunct}\relax
\EndOfBibitem
\bibitem[Ruffieux \emph{et~al.}(2016)Ruffieux, Wang, Yang, Sanchez-Sanchez,
  Liu, Dienel, Talirz, Shinde, Pignedoli, Passerone, Dumslaff, Feng, M\"ollen,
  and Fasel]{Ruffieux2016489}
P.~Ruffieux, S.~Wang, B.~Yang, C.~Sanchez-Sanchez, J.~Liu, T.~Dienel,
  L.~Talirz, P.~Shinde, C.~Pignedoli, D.~Passerone, T.~Dumslaff, X.~Feng,
  K.~M\"ollen and R.~Fasel, \emph{Nature}, 2016, \textbf{531}, 489--492\relax
\mciteBstWouldAddEndPuncttrue
\mciteSetBstMidEndSepPunct{\mcitedefaultmidpunct}
{\mcitedefaultendpunct}{\mcitedefaultseppunct}\relax
\EndOfBibitem
\bibitem[Steiner \emph{et~al.}(2017)Steiner, Gebhardt, Ammon, Yang,
  Heidenreich, Hammer, G{\"o}rling, Kivala, and Maier]{steiner2017hierarchical}
C.~Steiner, J.~Gebhardt, M.~Ammon, Z.~Yang, A.~Heidenreich, N.~Hammer,
  A.~G{\"o}rling, M.~Kivala and S.~Maier, \emph{Nat. Commun.}, 2017,
  \textbf{8}, 1--11\relax
\mciteBstWouldAddEndPuncttrue
\mciteSetBstMidEndSepPunct{\mcitedefaultmidpunct}
{\mcitedefaultendpunct}{\mcitedefaultseppunct}\relax
\EndOfBibitem
\bibitem[Clair and de~Oteyza(2019)]{Clair2019-OSS}
S.~Clair and D.~G. de~Oteyza, \emph{Chemical Reviews}, 2019, \textbf{119},
  4717--4776\relax
\mciteBstWouldAddEndPuncttrue
\mciteSetBstMidEndSepPunct{\mcitedefaultmidpunct}
{\mcitedefaultendpunct}{\mcitedefaultseppunct}\relax
\EndOfBibitem
\bibitem[Shu \emph{et~al.}({2018})Shu, Liu, Zha, Pan, Zhang, Xie, Chen, Yuan,
  Qiu, and Liu]{Shu_NatComm_2018}
C.-H. Shu, M.-X. Liu, Z.-Q. Zha, J.-L. Pan, S.-Z. Zhang, Y.-L. Xie, J.-L. Chen,
  D.-W. Yuan, X.-H. Qiu and P.-N. Liu, \emph{{Nat. Commun.}}, {2018},
  \textbf{{9}}, 2322\relax
\mciteBstWouldAddEndPuncttrue
\mciteSetBstMidEndSepPunct{\mcitedefaultmidpunct}
{\mcitedefaultendpunct}{\mcitedefaultseppunct}\relax
\EndOfBibitem
\bibitem[Zuzak \emph{et~al.}(2020)Zuzak, Jančařík, Gourdon, Szymonski, and
  Godlewski]{zuzak2020}
R.~Zuzak, A.~Jančařík, A.~Gourdon, M.~Szymonski and S.~Godlewski, \emph{ACS
  Nano}, 2020, \textbf{14}, 13316--13323\relax
\mciteBstWouldAddEndPuncttrue
\mciteSetBstMidEndSepPunct{\mcitedefaultmidpunct}
{\mcitedefaultendpunct}{\mcitedefaultseppunct}\relax
\EndOfBibitem
\bibitem[Galeotti \emph{et~al.}(2020)Galeotti, Marchi, Hamzehpoor, MacLean,
  Rao, Chen, Besteiro, Dettmann, Ferrari, Frezza, Sheverdyaeva, Liu, Kundu,
  Moras, Ebrahimi, Gallagher, Rosei, Perepichka, and
  Contini]{Galeotti-NatMat2020}
G.~Galeotti, F.~D. Marchi, E.~Hamzehpoor, O.~MacLean, M.~R. Rao, Y.~Chen,
  L.~Besteiro, D.~Dettmann, L.~Ferrari, F.~Frezza, P.~Sheverdyaeva, R.~Liu,
  A.~Kundu, P.~Moras, M.~Ebrahimi, M.~Gallagher, F.~Rosei, D.~Perepichka and
  G.~Contini, \emph{Nat. Mater.}, 2020, \textbf{19}, 874–880\relax
\mciteBstWouldAddEndPuncttrue
\mciteSetBstMidEndSepPunct{\mcitedefaultmidpunct}
{\mcitedefaultendpunct}{\mcitedefaultseppunct}\relax
\EndOfBibitem
\bibitem[Basagni \emph{et~al.}(2015)Basagni, Sedona, Pignedoli, Cattelan,
  Nicolas, Casarin, and Sambi]{Sedona}
A.~Basagni, F.~Sedona, C.~A. Pignedoli, M.~Cattelan, L.~Nicolas, M.~Casarin and
  M.~Sambi, \emph{J. Am. Chem. Soc.}, 2015, \textbf{137}, 1802--1808\relax
\mciteBstWouldAddEndPuncttrue
\mciteSetBstMidEndSepPunct{\mcitedefaultmidpunct}
{\mcitedefaultendpunct}{\mcitedefaultseppunct}\relax
\EndOfBibitem
\bibitem[Sun \emph{et~al.}(2017)Sun, Tran, Cai, Ma, Yu, Yuan, Stöhr, and
  Xu]{Sun_dehalogenative}
Q.~Sun, B.~V. Tran, L.~Cai, H.~Ma, X.~Yu, C.~Yuan, M.~Stöhr and W.~Xu,
  \emph{Angew. Chem.}, 2017, \textbf{129}, 12333--12337\relax
\mciteBstWouldAddEndPuncttrue
\mciteSetBstMidEndSepPunct{\mcitedefaultmidpunct}
{\mcitedefaultendpunct}{\mcitedefaultseppunct}\relax
\EndOfBibitem
\bibitem[Yu \emph{et~al.}(2020)Yu, Cai, Bao, Sun, Ma, Yuan, and Xu]{Yu2020}
X.~Yu, L.~Cai, M.~Bao, Q.~Sun, H.~Ma, C.~Yuan and W.~Xu, \emph{Chem. Commun.},
  2020, \textbf{56}, 1685--1688\relax
\mciteBstWouldAddEndPuncttrue
\mciteSetBstMidEndSepPunct{\mcitedefaultmidpunct}
{\mcitedefaultendpunct}{\mcitedefaultseppunct}\relax
\EndOfBibitem
\bibitem[Rabia \emph{et~al.}(2019)Rabia, Tumino, Milani, Russo, {Li Bassi},
  Achilli, Fratesi, Onida, Manini, Sun,\emph{et~al.}]{rabia2019scanning}
A.~Rabia, F.~Tumino, A.~Milani, V.~Russo, A.~{Li Bassi}, S.~Achilli,
  G.~Fratesi, G.~Onida, N.~Manini, Q.~Sun \emph{et~al.}, \emph{Nanoscale},
  2019, \textbf{{11}}, {18191--18200}\relax
\mciteBstWouldAddEndPuncttrue
\mciteSetBstMidEndSepPunct{\mcitedefaultmidpunct}
{\mcitedefaultendpunct}{\mcitedefaultseppunct}\relax
\EndOfBibitem
\bibitem[Rabia \emph{et~al.}(2020)Rabia, Tumino, Milani, Russo, Bassi, Bassi,
  Lucotti, Achilli, Fratesi, Manini, Onida, Sun, Xu, and Casari]{Rabia_nanomat}
A.~Rabia, F.~Tumino, A.~Milani, V.~Russo, A.~L. Bassi, N.~Bassi, A.~Lucotti,
  S.~Achilli, G.~Fratesi, N.~Manini, G.~Onida, Q.~Sun, W.~Xu and C.~S. Casari,
  \emph{ACS App. Nano Mater.}, 2020, \textbf{3}, 12178--12187\relax
\mciteBstWouldAddEndPuncttrue
\mciteSetBstMidEndSepPunct{\mcitedefaultmidpunct}
{\mcitedefaultendpunct}{\mcitedefaultseppunct}\relax
\EndOfBibitem
\bibitem[Achilli \emph{et~al.}(2021)Achilli, Milani, Fratesi, Tumino, Manini,
  Onida, and Casari]{Achilli-GDY}
S.~Achilli, A.~Milani, G.~Fratesi, F.~Tumino, N.~Manini, G.~Onida and C.~S.
  Casari, \emph{2D Mater}, 2021, \textbf{8}, 044014\relax
\mciteBstWouldAddEndPuncttrue
\mciteSetBstMidEndSepPunct{\mcitedefaultmidpunct}
{\mcitedefaultendpunct}{\mcitedefaultseppunct}\relax
\EndOfBibitem
\bibitem[Xu \emph{et~al.}({2017})Xu, Kojima, Wang, Kaushik, Saliniemi, Nakae,
  and Sakaguchi]{Xu17}
Z.~Xu, T.~Kojima, W.~Wang, K.~Kaushik, A.~Saliniemi, T.~Nakae and H.~Sakaguchi,
  \emph{{Mater. Chem. Front.}}, {2017}, \textbf{{2}}, {775--779}\relax
\mciteBstWouldAddEndPuncttrue
\mciteSetBstMidEndSepPunct{\mcitedefaultmidpunct}
{\mcitedefaultendpunct}{\mcitedefaultseppunct}\relax
\EndOfBibitem
\bibitem[Wang \emph{et~al.}({2021})Wang, Li, Ding, Mattioli, Huang, Wang,
  Gourdon, Sun, Chen, Kantorovich, Yang, Rosei, and Yu]{Wang21}
S.~Wang, Z.~Li, P.~Ding, C.~Mattioli, W.~Huang, Y.~Wang, A.~Gourdon, Y.~Sun,
  M.~Chen, L.~Kantorovich, K.~Yang, F.~Rosei and M.~Yu, \emph{{Angew. Chem}},
  {2021}, \textbf{{60}}, {17435--17439}\relax
\mciteBstWouldAddEndPuncttrue
\mciteSetBstMidEndSepPunct{\mcitedefaultmidpunct}
{\mcitedefaultendpunct}{\mcitedefaultseppunct}\relax
\EndOfBibitem
\bibitem[Clair and de~Oteyza(2019)]{Chem-Rev}
S.~Clair and D.~G. de~Oteyza, \emph{Chem. Rev.}, 2019, \textbf{119},
  4717--4776\relax
\mciteBstWouldAddEndPuncttrue
\mciteSetBstMidEndSepPunct{\mcitedefaultmidpunct}
{\mcitedefaultendpunct}{\mcitedefaultseppunct}\relax
\EndOfBibitem
\bibitem[Wang and Zhu(2019)]{Rev-SSR}
T.~Wang and J.~Zhu, \emph{Surf. Sci. Rep.}, 2019, \textbf{74}, 97--240\relax
\mciteBstWouldAddEndPuncttrue
\mciteSetBstMidEndSepPunct{\mcitedefaultmidpunct}
{\mcitedefaultendpunct}{\mcitedefaultseppunct}\relax
\EndOfBibitem
\bibitem[Xiang \emph{et~al.}({2019})Xiang, Lu, Wang, Ju, Di~Filippo, Li, Liu,
  Leng, Zhu, Wang, and Schneider]{Xiang19}
F.~Xiang, Y.~Lu, Z.~Wang, H.~Ju, G.~Di~Filippo, C.~Li, X.~Liu, X.~Leng, J.~Zhu,
  L.~Wang and M.~A. Schneider, \emph{{J. Phys. Chem. C}}, {2019},
  \textbf{{123}}, {23007–23013}\relax
\mciteBstWouldAddEndPuncttrue
\mciteSetBstMidEndSepPunct{\mcitedefaultmidpunct}
{\mcitedefaultendpunct}{\mcitedefaultseppunct}\relax
\EndOfBibitem
\bibitem[Cirera \emph{et~al.}(2020)Cirera, S\'anchez-{G}rande, de~la {T}orre,
  Santos, Edalatmanesh, Rodr\'iguez-{S}\'anchez, Lauwaet, Mallada, Zbořil,
  Miranda, Gr\"oning, Jel\'inek, Mart\'in, and Ecija]{Cirera}
B.~Cirera, A.~S\'anchez-{G}rande, B.~de~la {T}orre, J.~Santos, S.~Edalatmanesh,
  E.~Rodr\'iguez-{S}\'anchez, K.~Lauwaet, B.~Mallada, R.~Zbořil, R.~Miranda,
  O.~Gr\"oning, P.~Jel\'inek, N.~Mart\'in and D.~Ecija, \emph{Nat.
  Nanotechnol.}, 2020, \textbf{15}, 437–443\relax
\mciteBstWouldAddEndPuncttrue
\mciteSetBstMidEndSepPunct{\mcitedefaultmidpunct}
{\mcitedefaultendpunct}{\mcitedefaultseppunct}\relax
\EndOfBibitem
\bibitem[Kawai \emph{et~al.}({2018})Kawai, Krejčí, Foster, Pawlak, Xu, Peng,
  Orita, and Meyer]{Kawai18}
S.~Kawai, O.~Krejčí, A.~S. Foster, R.~Pawlak, F.~Xu, L.~Peng, A.~Orita and
  E.~Meyer, \emph{{ACS Nano}}, {2018}, \textbf{{12}}, {8791--8796}\relax
\mciteBstWouldAddEndPuncttrue
\mciteSetBstMidEndSepPunct{\mcitedefaultmidpunct}
{\mcitedefaultendpunct}{\mcitedefaultseppunct}\relax
\EndOfBibitem
\bibitem[Fratesi \emph{et~al.}(2018)Fratesi, Achilli, Manini, Onida, Baby,
  Ravikumar, Ugolotti, Brivio, Milani, and Casari]{ma11122556}
G.~Fratesi, S.~Achilli, N.~Manini, G.~Onida, A.~Baby, A.~Ravikumar,
  A.~Ugolotti, G.~P. Brivio, A.~Milani and C.~S. Casari, \emph{Materials},
  2018, \textbf{11}, 2556\relax
\mciteBstWouldAddEndPuncttrue
\mciteSetBstMidEndSepPunct{\mcitedefaultmidpunct}
{\mcitedefaultendpunct}{\mcitedefaultseppunct}\relax
\EndOfBibitem
\bibitem[Goudappagouda \emph{et~al.}(2017)Goudappagouda, Wakchaure, Ranjeesh,
  Abhai, and Babu]{synthexp}
Goudappagouda, V.~C. Wakchaure, K.~C. Ranjeesh, C.~A.~R. Abhai and S.~S. Babu,
  \emph{Chem. Commun.}, 2017, \textbf{53}, 7072--7075\relax
\mciteBstWouldAddEndPuncttrue
\mciteSetBstMidEndSepPunct{\mcitedefaultmidpunct}
{\mcitedefaultendpunct}{\mcitedefaultseppunct}\relax
\EndOfBibitem
\bibitem[Perdew \emph{et~al.}(1996)Perdew, Burke, and Ernzerhof]{PBE}
J.~P. Perdew, K.~Burke and M.~Ernzerhof, \emph{Phys. Rev. Lett.}, 1996,
  \textbf{77}, 3865--3868\relax
\mciteBstWouldAddEndPuncttrue
\mciteSetBstMidEndSepPunct{\mcitedefaultmidpunct}
{\mcitedefaultendpunct}{\mcitedefaultseppunct}\relax
\EndOfBibitem
\bibitem[Grimme(2006)]{Grimme}
S.~Grimme, \emph{J. Comput. Chem.}, 2006, \textbf{27}, 1787--1799\relax
\mciteBstWouldAddEndPuncttrue
\mciteSetBstMidEndSepPunct{\mcitedefaultmidpunct}
{\mcitedefaultendpunct}{\mcitedefaultseppunct}\relax
\EndOfBibitem
\bibitem[Soler \emph{et~al.}(2002)Soler, Artacho, Gale, Garc{\'\i}a, Junquera,
  Ordej\'on, and S\'anchez-Portal]{Sole02}
J.~M. Soler, E.~Artacho, J.~D. Gale, A.~Garc{\'\i}a, J.~Junquera, P.~Ordej\'on
  and D.~S\'anchez-Portal, \emph{J. Phys.: Condens. Matter}, 2002, \textbf{14},
  2745\relax
\mciteBstWouldAddEndPuncttrue
\mciteSetBstMidEndSepPunct{\mcitedefaultmidpunct}
{\mcitedefaultendpunct}{\mcitedefaultseppunct}\relax
\EndOfBibitem
\bibitem[Tersoff and Hamman(1985)]{Tersoff}
J.~Tersoff and D.~R. Hamman, \emph{Phys. Rev. B}, 1985, \textbf{31}, 805\relax
\mciteBstWouldAddEndPuncttrue
\mciteSetBstMidEndSepPunct{\mcitedefaultmidpunct}
{\mcitedefaultendpunct}{\mcitedefaultseppunct}\relax
\EndOfBibitem
\end{mcitethebibliography}
\bibliographystyle{rsc} %the RSC's .bst file

\end{document}